\documentclass{elsarticle}

\usepackage{amssymb,amsmath,framed,verbatim,tikz}
\usetikzlibrary{shapes}
\usepackage{latexsym}
\newtheorem{theorem}{Theorem}[section]
\newtheorem{lemma}[theorem]{Lemma}
\newtheorem{proposition}[theorem]{Proposition}
\newtheorem{claim}[theorem]{Claim}

\newtheorem{corollary}[theorem]{Corollary}

\newtheorem{remark}[theorem]{Remark}
\newtheorem{observation}[theorem]{Observation}

\def\l{{\lambda}}
\def\d{{\delta}}

\def\D{{\rm {\bf D}}}
\def\E{{\rm {\bf E}}}
\def\Var{{\rm {\bf Var}}}
\def\SR{{\mathbb{R}}}

\def\P{{\cal P}}

\def\r{\rho}
\def\qedf{\hfill $\Box$}

\newcommand{\mat}[2]{\left[\begin{array}{#1}#2\end{array}\right]}

\date{}

\title{{Optimal Random Matchings, Tours, and Spanning Trees \\ in Hierarchically Separated Trees}$^1$}
\author{{\large B\'ela Csaba$^{a,2}$, Thomas A. Plick$^{b,3}$, and Ali Shokoufandeh$^{b,4}$}
{\small \shortstack[l]{ \\ \\
$^a$ Department of Mathematics, Western Kentucky University, \\
\hspace{1in}    1906 College Heights Boulevard, Bowling Green KY 42101, USA \\
$^b$ Department of Computer Science, Drexel University, \\
\hspace{1in}    3141 Chestnut Street, Philadelphia PA 19104, USA
}}}

\begin{document}


\begin{abstract}
We derive tight bounds on the expected
weights of several combinatorial optimization problems  
for random point sets of size $n$ distributed among
the leaves of a balanced hierarchically separated tree.
We consider {\it monochromatic} and
{\it bichromatic} versions of the minimum
matching, minimum spanning tree, and
traveling salesman problems.
We also present tight concentration results for the monochromatic problems.

\end{abstract}

\maketitle

\footnotetext[1]{The authors were partially supported by
National Science Foundation grant \#0803670 under the IIS Division and Office of Naval Research grant
ONRN000140410363.}
\footnotetext[2]{This author was partially supported by the Hungarian Scientific Research Fund (OTKA) grant K76099 and by ERC, grant no.~258345.  He is also supported by the European Union and co-funded by the European Social Fund
 under the project ``Telemedicine-focused research activities on the field of Mathematics,
 Informatics and Medical sciences'' of project number ``T\'AMOP-4.2.2.A-11/1/KONV-2012-0073''.
He has since moved to the University of Szeged, where his address is as follows:
B\'ela Csaba, Bolyai Institute,
University of Szeged, Aradi v\'ertan\'uk tere 1, Szeged, 6724 Hungary.
His new email address is {\tt bcsaba@math.u-szeged.hu}}
\footnotetext[3]{Corresponding author.  Email address: {\tt tap42@drexel.edu}; phone: 215-895-2669;
 fax: 215-895-0545}
\footnotetext[4]{Email address: {\tt ashokouf@cs.drexel.edu}}

\section{Introduction}

The problem of computing a large similar common subset of two point
sets arises in many areas of computer science, from computer vision
and pattern recognition to bio-informatics~\cite{Rubner00}. Most of
the recent related work concerns the design of efficient algorithms to
compute rigid transformations for establishing correspondences between
two point sets in $\SR^d$ subject to minimization of a distance
measure. Comparatively little attention has been devoted to extremal
matching problems related to random point sets, such as the following problem:
{\em Presented with two random point sets, how do we expect matching weight to vary
  with data set size?}

The seminal work in extremal random matching is, arguably, the
1984 paper of Ajtai, Koml\'os and Tusn\'ady~\cite{Ajtai84}, whose
deep result found many applications in the following
years.  For two infinite point sets $X$ and $Y$ chosen
independently and uniformly at random from $[0, 1]^2$, they derived asymptotic bounds
on the sequence $\{\E M_n\}_{n=1}^\infty$, where $M_i$ is the optimal matching weight
between $\{x_1, \ldots, x_n\}$ and $\{y_1, \ldots, y_n\}$.
A short time later, Leighton and Shor~\cite{Leighton86} addressed the
related problem of determining the expectation of the \emph{maximum}
cost of any edge in the matching (instead of the sum).
Shor~\cite{Shor86} subsequently applied the AKT result to obtain
bounds on the average-case analysis of several
algorithms. Talagrand~\cite{Talagrand96} introduced the notion of
majorizing measures and, as an illustration of this powerful
technique, derived the theorem of Ajtai et al. Rhee and
Talagrand~\cite{Rhee88} explored the two-dimensional \emph{upward matching} problem,
in which points from $X$ must be matched to points of
$Y$ that have greater $x$- and $y$-coordinates.  They have also
explored a similar problem in the
cube~\cite{Rhee92}. In~\cite{Talagrand92} Talagrand gave insight into
the exact behavior of expected matching weight for dimensions $d\ge 3$
under arbitrary norms.

Our main goal in this paper is to continue the investigations started in
\cite{AbrahamsonPhD} and \cite{Abrahamson} into the probability
theory of some classical combinatorial optimization problems on
hierarchically separated trees (HSTs).  The notion of the
hierarchically (well-)separated tree was introduced by
Bartal~\cite{Bartal96}.
A $\l$-HST is a rooted weighted tree with two
additional properties: (1) edge weights between nodes and their
children are the same for any given parent node, and (2) the ratio of
incident edge weights along a root-leaf path is $\l$ (so edges become
lighter by a factor of $\l$ as one approaches leaves).  In what
follows, we only consider HSTs that are balanced (those for which the
branching factor of all nodes other than the leaves is the same,
denoted by $b$) and uniform (having every leaf at the same depth
$\d$).  We also require $0 < \lambda < 1$.  For the sake of brevity,
we describe an HST with these parameters as a $(b, \delta, \lambda)$-HST.

Bartal showed \cite{Bartal98} that arbitrary metric spaces on $n$ points can be $O(\log n \log \log n)$-probabilistically approximated
by a collection of HSTs; the factor of the approximation (here $c\log n \log \log n$)
indicates the maximum factor by which a randomly chosen tree from the collection will overestimate, in expectation,
the distance between any pair of points.
Fakcharoenphol, Rao and Talwar~\cite{Fakcharoenphol03} improved Bartal's result to apply to $O(\log n)$-probabilistically approximated
by a collection of HSTs. This implies that our bounds for HSTs translate to other metric spaces.

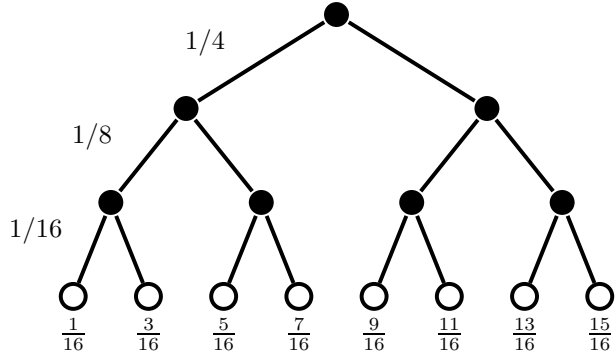
\begin{figure}
\begin{center}
\begin{tikzpicture}[scale=2.5,line width=1.5pt]
    \node (root) at (0, 0) [circle,fill] {};
    
    \node (11) at (-0.8, -0.5) [circle,fill] {};
    \node (12) at ( 0.8, -0.5) [circle,fill] {};

    \node (21) at (-1.2, -1) [circle,fill] {};
    \node (22) at (-0.4, -1) [circle,fill] {};
    \node (23) at ( 0.4, -1) [circle,fill] {};
    \node (24) at ( 1.2, -1) [circle,fill] {};
    
    \node (31) at (-1.4, -1.5) [circle, draw] {};
    \node (32) at (-1.0, -1.5) [circle, draw] {};
    \node (33) at (-0.6, -1.5) [circle, draw] {};
    \node (34) at (-0.2, -1.5) [circle, draw] {};
    \node (35) at ( 0.2, -1.5) [circle, draw] {};
    \node (36) at ( 0.6, -1.5) [circle, draw] {};
    \node (37) at ( 1.0, -1.5) [circle, draw] {};
    \node (38) at ( 1.4, -1.5) [circle, draw] {};
    
    \node at (-1.4, -1.7) {$\frac{1}{16}$};
    \node at (-1.0, -1.7) {$\frac{3}{16}$};
    \node at (-0.6, -1.7) {$\frac{5}{16}$};
    \node at (-0.2, -1.7) {$\frac{7}{16}$};
    \node at ( 0.2, -1.7) {$\frac{9}{16}$};
    \node at ( 0.6, -1.7) {$\frac{11}{16}$};
    \node at ( 1.0, -1.7) {$\frac{13}{16}$};
    \node at ( 1.4, -1.7) {$\frac{15}{16}$};
    
    \node at (-0.7, -0.15) {1/4};
    \node at (-1.3, -0.65) {1/8};
    \node at (-1.6, -1.15) {1/16};
    
    \draw (root) -- (11);
    \draw (root) -- (12);
    
    \draw (11) -- (21);
    \draw (11) -- (22);
    \draw (12) -- (23);
    \draw (12) -- (24);
    
    \draw (21) -- (31);
    \draw (21) -- (32);
    \draw (22) -- (33);
    \draw (22) -- (34);
    \draw (23) -- (35);
    \draw (23) -- (36);
    \draw (24) -- (37);
    \draw (24) -- (38);
\end{tikzpicture}

\caption{
    Eight equispaced points on $[0,1]$ and their HST approximation.
    For every pair of points $x, y$, the distance between $x$ and $y$
    in the tree exceeds or equals their Euclidean distance $|x-y|$.
    \label{fig:firsthst}
}
\end{center}
\end{figure}

General HSTs are well suited to approximating arbitrary Euclidean metric spaces.
In this paper we consider only balanced uniform HSTs, but even these are useful in approximating ``well-behaved''
spaces.  To start, let us consider a set $P_n$ of $n$ equally spaced points on the interval $[0,1]$.  For simplicity,
let us assume the points are placed at the midpoints of the subintervals $[0,1/n], [1/n,2/n], \ldots, [(n-1)/n, 1]$.
For $n$ a power of two, we can draw a full binary tree of height $\log_2 n$ whose leaves are collocated with $P_n$.
Figure \ref{fig:firsthst} shows the tree for $n=8$.  The edge weight ratio is set to $1/2$, and the edges incident on the root are
given weights of $1/4$; the tree then has a diameter of $1-1/n$.

This tree defines a metric on the leaves of $P_n$:
the distance between two leaves is the length of the path connecting them.
It is clear that the distance between the two leaves $p$ and $q$ in the tree is greater than or equal to
the Euclidean distance between the points $p$ and $q$; thus it is said that the tree metric \emph{dominates}
the Euclidean metric on these points.  While the tree metric is close to the Euclidean distance for many point pairs
(consider the two extreme points), other point pairs are assigned an exaggerated distance.  Consider
the two middle points: their Euclidean distance is $1/n$, but their distance in the tree is nearly $1$.

\begin{figure}
\begin{center}
\begin{tikzpicture}[scale=2.5,line width=1.5pt]
\input{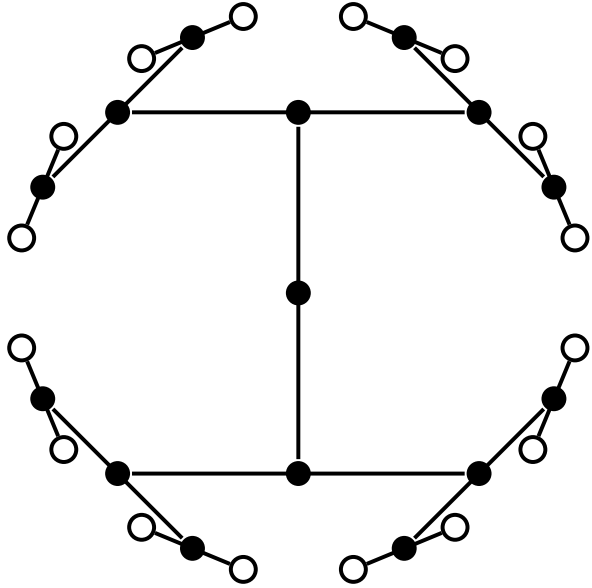}
\end{tikzpicture}

\caption{
    The sixteen white points represent sixteen equispaced points on the toric interval $[0, 2]$.
    They are paired off, and the midpoint of each pair is drawn, with edges to the points
    of its pair.  This process is repeated until only one point remains.    
    \label{fig:secondhst}
}
\end{center}
\end{figure}

We obtain results with a more complex technique.  Let us embed $P_n$ into the interval $[0,2]$ equipped with
a torus distance metric: the distance between two points $x$ and $y$ is defined as $\min(|x-y|, 2-|x-y|)$.
Thus the interval wraps around --- the points 0 and 2 are coincident --- but for two points inside $[0,1]$,
the torus metric agrees with the Euclidean metric.  Define $P'_n = P_n \cup (1 + P_n)$.
We can repeat our construction from above: choose one of the points to be the ``first'' leaf,
and build a binary tree whose leaves are collocated with $P'_n$.  In this way we obtain a set of $n$ trees; one such tree
is shown in Figure \ref{fig:secondhst}.  Now,
for a fixed pair of points $p$ and $q$ from $P_n$, some trees will greatly overestimate the distance between $p$ and $q$,
but most of the trees will give a fairly close value to the Euclidean distance.  Choosing a tree at random
will provide an approximation that is accurate with high probability; one can show that the expected
distance in the tree is no more than $(\log_2 n)$ times the actual Euclidean distance.  A tight bound on tree distances,
therefore, translates into a bound on the Euclidean distances that is tight to within a factor of $\log_2 n$.

Our construction easily generalizes to higher dimensions.  Given $n$ points equally spaced in $[0,1]^d$,
we can embed them into a set of $n\cdot 2^d$ points in $[0,2]^d$ equipped with a torus metric.  The resulting tree will now have a branching
factor of $2^d$.
Furthermore, the restriction to equispaced points is, for many applications, not essential: points
on $[0, 1]^d$ can be discretized to an equispaced grid of points.  The most convenient such grid
is the one formed by cutting the hypercube into $n$ parts along every dimension, forming $n^d$ smaller hypercubes,
and taking the midpoint of each cube.
This setup introduces some error, which depends on the coarseness of the discretization;
the optimal error is often negligible when compared to the cost of the problem.
For more on the application of HSTs to problems on metric spaces, we refer the reader
to the papers by Bartal \cite{Bartal96,Bartal98} and Fakcharoenphol et al.~\cite{Fakcharoenphol03}.
The method of successive grid approximations and its utility are examined in greater depth
in Indyk and Thaper's paper \cite{Indyk} on applying the earth-mover distance to fast image
retrieval.


Clearly, every HST determines a metric space on its leaves. 
In the present paper we study some combinatorial optimization problems on metric spaces of balanced HSTs. We extend the results of
Abrahamson~\cite{AbrahamsonPhD} and Abrahamson et
al.~\cite{Abrahamson}, which considered the bichromatic random
matching problem on HSTs.  They showed
\begin{theorem}
\label{bichromatic_matching}[1, 2]
Let $T=T(b,\d,\l)$ be a balanced HST; let $R$ and $B$ be two randomly
chosen $n$-element submultisets of the set of leaves of $T$; and define
$h=\min{(\d, \log_b{n})}.$ Then there exist positive constants $K_1$ and
$K_2$ such that
$$K_1 \sqrt{b n}\sum_{k=1}^{h}(\sqrt{b}\l)^k \le \E M_T(R,B) \le 
K_2 \sqrt{b n}\sum_{k=1}^{h}(\sqrt{b}\l)^k.$$
\end{theorem}
We will presently add to these results the cases of (1) monochromatic
matching on HSTs, (2) monochromatic and bichromatic TSP, and (3)
monochromatic and bichromatic MSTs.
In all these problems we look at an edge-weighted graph $G$ that is induced by
the randomly chosen leaves of the HST. The edge weight of two points will
be their distance in the tree.
We will estimate the expected total weight of some subgraphs of $G.$




The remainder of this paper is laid out as follows.
Section \ref{defs} contains the formal definitions of the problems
we consider.
Section \ref{monomatching} gives the bound on monochromatic matching.
We consider the monochromatic TSP and MST problem in Section \ref{mono_tsp};
the bichromatic MST problem in Section \ref{bi_mst};
and the bichromatic TSP in Section \ref{bi_tsp}.
In Section~\ref{concentration} we demonstrate the very tight concentration of the
cost of our monochromatic optimization problems around their mean.
In Section \ref{conclusion} we offer our conclusions and ideas for future work.

\section{Definitions of problems}\label{defs}

Given an HST $H$ and a submultiset $V'$ of its leaves, we let $G$
be the complete graph with vertex set $V.$ Here $V$ contains the elements of $V',$ and if there are multiple copies of a leaf in $V',$
we distinguish them in $V$ by indexing the different copies. For every $v, w\in V$ the
weight of the path from $v$ to $w$ is denoted by $d_H(v, w).$ 
Note that we will have distinct
vertices of $G$ that are at a distance zero from each other, when
these are copies of the same leaf of $H.$

First we define the monochromatic versions of the problems we will
consider.  (Our definitions follow Yukich \cite{Yukich}.)

{\bf Minimum Matching Problem ($MMP$):} The minimum
  matching on $V$ has cost given
  by 
  $$M(V)=\min_{M\subset G}\sum_{e \in M} |e|,$$
  where $|e|$ denotes the weight of the
  edge $e$ and the minimum is taken over all possible perfect matchings of $G.$
  If $n$ is odd, one of the vertices is excluded from the matching, and
  so the minimum matching on $V$ is the minimum of the minimum
  matchings on the $n$ subsets $V \backslash \{x_1\}, V \backslash
  \{x_2\}, \ldots, V \backslash \{x_n\}$.

{\bf Traveling Salesman Problem ($TSP$):} A closed tour or
  Hamiltonian cycle is a simple cycle that visits each vertex of $V$
  exactly once. We denote by $T(V)$ the length of the
  shortest closed tour $T$ on $V$.
  Thus $$T(V)=\min_{T\subset G}\sum_{e \in T}|e|,$$ where the minimum is
  taken over all tours $T$.

{\bf Minimum Spanning Tree Problem ($MST$):}
  Let $MST(V)$ be the cost of the shortest spanning tree
  on $V$,
  namely $$MST(V)=\min_{F\subset G}\sum_{e\in F}|e|,$$ where the minimum is
  taken over all spanning trees $F$.


In the bichromatic versions of the above problems we first choose two submultisets of the leaves of $H$; let us call these $R'$ and $B'.$ Then $G$ will be the complete bipartite
graph with two $n$-element color classes $R$ and $B,$ where, as before, $R$ will contain every element of $R',$ $B$ will contain every element of $B',$ and, if necessary, we index
the multiple copies of the leaves. The edge weights again are determined by the distance of the endpoints in $H.$ 

The bichromatic versions of the above problems are defined very similarly to the monochromatic cases.
Since $G$ is bipartite, every edge that will be included
in the solution of any of the optimization problems above will be a
red-blue edge: the set of matchings $M$, the set of tours $T$, and the set of spanning trees $F$
are now restricted to those subgraphs in which each edge is bichromatic.  For completeness we list these
problems below.

{\bf Bichromatic Minimum Matching Problem ($MMP'$):} The minimum
  matching on $V$ has cost given
  by 
  $$\min_{M\subset G}\sum_{e \in M} |e|,$$
  where $|e|$ denotes the weight of the
  edge $e$ and the minimum is taken over all possible perfect bichromatic matchings of $G.$
  If $n$ is odd, one of the vertices is excluded from the matching, and
  so the minimum matching on $V$ is the minimum of the minimum bichromatic
  matchings on the $n$ subsets $V \backslash \{x_1\}, V \backslash
  \{x_2\}, \ldots, V \backslash \{x_n\}$.

{\bf Bichromatic Traveling Salesman Problem ($TSP'$):} A bichromatic tour or
  is a simple cycle that visits each vertex of $V$
  exactly once and alternates colors along each edge. The minimum tour length is given by
  $$\min_{T\subset G}\sum_{e \in T}|e|,$$ where the minimum is
  taken over all bichromatic tours $T$.

{\bf Bichromatic Minimum Spanning Tree Problem ($MST'$):}
  The minimum bichromatic spanning tree cost
  is given by $$\min_{F\subset G}\sum_{e\in F}|e|,$$ where the minimum is
  taken over all bichromatic spanning trees $F$.


\section{Monochromatic matching}\label{monomatching}

Let $H$ be an HST and $W$ be a set of points assigned to the leaves of $H$.  For each vertex $v \in V(H)$,
we define $X(v)$ to be the number of points assigned to the descendants of $v$ in $H$.  
We consider a node to be a descendant of itself, so that when $v$ is a leaf, $X(v)$ is simply the number
of points assigned to $v$.

For a non-leaf vertex $v$ with
children $u_1, u_2, \ldots, u_b$, we have \[X(v) = \sum_{i=1}^b X(u_i).\]


For a pair of matched points $(x, y)$ in $M(W)$, belonging to distinct
leaves $u_x$ and $u_y$ in $H$, we will say the pair $(x, y)$
results in a {\em transit} at vertex $v$ when $v$ is the lowest common
ancestor of $u_x$ and $u_y$, that is, the path between $u_x$ and $u_y$
passes through $v$.  We will also use $\tau_v$ to denote the total
number of transits at vertex $v$ in an optimal matching. Given an
integer $N$ we define $Odd(N)$ to be the parity of $N$, i.e. 1 when $N$ is odd
and 0 when $N$ is even.

\begin{lemma}\label{transit}
Let $H$ be an HST, and let $W$ be a submultiset of the leaves of $H.$
The number of transits at any non-leaf vertex $v$ in a minimum
matching $M(W)$ is $$\tau_v=\frac{1}{2}\left(\left(\sum_{i=1}^b
Odd(X(u_i))\right)-Odd\left(X(v)\right)\right),$$ where $u_1, u_2,
\ldots, u_b$ are the children of $v$.
\end{lemma}


\noindent {\bf Proof:} First consider a vertex $v$ at height 1 and its children
$u_1, u_2, \ldots, u_b$.  For each $u_i$, only $Odd(X(u_i))$ instances
need to look for a match elsewhere in the tree, since an even number
of the instances can be paired off in the leaf.  The number of
remaining unmatched points will therefore be $\sum_i Odd(X(u_i))$.
These points can now be paired off at $v$; if this quantity is odd,
there will be one point left over.  The number of pairs that transit
through $v$ is thus
\[\frac12\left(\sum_i Odd(X(u_i)) - Odd\left(\sum_i Odd(X(u_i))\right)\right) = \frac12\left(\sum_i Odd(X(u_i)) - Odd\left(\sum_i X(u_i)\right)\right).\]

We will now modify the HST in the following way.  Each point resident
at a leaf of the tree will be transferred to its parent, and the
leaves will be removed.  Thus each vertex $v$ that was formerly at
depth 1 will become a leaf, and every remaining vertex will retain its
previous $X$-value.  It is clear that the number of transits at every
vertex $w$ in this new tree remains the same, since any
unmatched point at $u_i$ would have had to search above $v$ in any
case, and these points will transit at the same vertex as they would
have before.
Successive applications of the argument in the preceding paragraph will reduce
the tree to a single node, showing that
$\tau_v=\frac{1}{2}\left(\sum_i Odd(X(u_i))-Odd\left(\sum_i
X(u_i)\right)\right)$ for every vertex $v$ in the original tree.
\hfill\qedf

The above argument also shows that $\tau_v \le \frac 1 2 deg_H(v)$: the number of transits at $v$ is upper-bounded by the number of children of $v$,
regardless of the number of descendants $v$ has in the HST.

\begin{lemma}\label{lemma:half_stronger}
Let $v$ be a non-root vertex in level $\ell$ of an HST $H$.
Then \[Prob[X(v) {\rm \,\, is \,\, odd}] \leq \frac 12,\]
so that $\E[Odd(X(v))] \leq \frac12$.
If $n$, the number of points chosen,
is at least $b^{\ell}$, then \[Prob[X(v) {\rm \,\, is \,\, odd}] \geq \frac 14,\]
so that $\E[Odd(X(v))] \geq \frac 14$.
\end{lemma}
{\bf Proof:} Denote the total number of leaves of $H$ by $s.$ Observe, that the number of leaves of $H$ that are descendant of $v$ is exactly $s/b^{\ell}.$ Set $r = 1/b^\ell$.
Let us start with an empty multiset of points and add
points one by one.  Some of these points will belong to leaves that
are descendants of $v$, while others will not.  For $i = 1, \ldots,
n$, define $m_i$ to be the number of the first $i$ points that belong
to descendants of $v$, and let $\chi_i = Odd(m_i)$.  The variables
$\chi_i$ form a Markov process where each transition changes state
with probability $r$ and remains at the current state with probability
$1-r$; this gives us the transition matrix $\P = \mat{cc}{1 - r & r
  \\ r & 1 - r}$.  The initial state of the process is $\pi = [1, \,\,
  0]$.

We can diagonalize $\P$ as $U^{-1}DU$, where $D = \mat{cc}{1 & 0 \\ 0
  & 1 - 2r}$ and $U = U^{-1} = \frac 1 {\sqrt 2} \mat{rr}{1 & 1 \\ 1 &
  -1}$.  After $n$ points have been chosen, the state of the system
will be $\pi \P^n = \pi U^{-1} D^n U = $ \\
$\left[\frac12 + \frac12 (1-2r)^n, \,\, \frac12 - \frac12 (1-2r)^n \right],$
and $Prob[X(v) {\rm \,\, is \,\, odd}] = \frac12 - \frac12 (1-2r)^n$.

Set $\Delta = (1-2r)^n$.  Notice that $\Delta$ cannot be negative,
since $b \geq 2$ and $r \leq 1/b$.  This shows the first statement of
the lemma.  For the lower bound, we wish to show $\Delta \leq
\frac12$.  When $n \geq b^\ell$, the worst case is clearly for $r =
1/{b^{\ell}}$ and $n = b^{\ell}$, which gives us \[\Delta \leq
(1-2/b^{\ell})^{b^{\ell}} < \exp(-2) < \frac12,\] and the second statement
is proved.  \hfill \qedf

(The proof actually gives a tighter lower bound of $1/2 - 1/(2e^2)$, but
$1/4$ is good enough for our purposes.)

We now have

\begin{corollary}\label{corr:expectation}
Let $H = H(b, \d, \l)$ be a balanced HST, and let $v$ be one of its
non-leaf vertices in level $\ell > 0$.  Then \[\E[\tau_v] \leq \frac
14 (b-1),\] and if $n \geq b^{\ell}$, \[\E[\tau_v] \geq \frac 18 (b-1).\]
\end{corollary}

{\bf Proof:} Take the expectation of $\tau_v$ as given by
Lemma~\ref{transit}.  The values of $\E[Odd(X(v))]$ and
$\E[Odd(X(u_i))]$ are bounded by Lemma~\ref{lemma:half_stronger}.
\hfill \qedf

\begin{theorem}\label{thm:sum}
Let $H = H(b, \d, \l)$ be a balanced HST, and let $W$ be a randomly
chosen $n$-element submultiset of its leaves.  The value of
$\E[M(W)]$, the expected weight of the optimal matching of $W$, obeys
\[\E[M(W)] \leq B_{b, \l} \sum_{k=1}^{\d} (b\lambda)^k\]
for some positive constant $B_{b, \l}$ depending only on $b$ and $\l$.
If $n \geq b^\d$, then
\[\E[M(W)] \geq A_{b, \l} \sum_{k=1}^{\d} (b\lambda)^k\]
for some positive constant $A_{b, \l}$ depending only on $b$ and $\l$.
\end{theorem}

{\bf Proof:} Consider the contribution of level $k$ (where $k > 0$) to
the value of $M(W)$.  Level $k$ contains $b^k$ vertices, and the
expected number of transits at each of these vertices is between
$\frac 18 (b-1)$ and $\frac 14 (b-1)$, by
Corollary~\ref{corr:expectation}.  The weight of a match at level $k$
is $2(\lambda^k + \lambda^{k+1} + \lambda^{k+2} + \ldots + \l^{\d})$; this is bounded from below by $2\lambda^k$ and from above by
$2\sum_{i=k}^\infty \lambda^i = 2\l^k/(1 - \l)$.  The total
contribution of levels $1, 2, \ldots, \d$ can therefore be bounded
from above by \[\sum_{k=1}^{\d} b^k \cdot \frac14 (b-1) \cdot
\frac{2\l^k}{1-\l} = \frac{b-1}{2(1-\l)} \sum_{k=1}^{\d} (b\l)^k.\]
and bounded from below, when $n \geq b^\d$, by \[\sum_{k=1}^{\d} b^k
\cdot \frac18 (b-1) \cdot 2\l^k = \frac{b-1}{4} \sum_{k=1}^{\d}
(b\l)^k.\] Choose $A_{b,\l} < (b-1)/4$ and $B_{b,\l} > (b-1)/2(1-\l)$.  We have not
counted the contribution of the root, but it is negligible, at most the constant $b.$
\qedf

\begin{remark}
Let us perform a walk on the tree, starting at the root, and visiting
each leaf in order from left to right.  This walk will use each edge
twice.  At level $\ell$ there are $b^\ell$ edges, and their cost is
$\l^\ell$.  Therefore the total cost will be \[2 \cdot
\sum_{\ell=1}^{\d} b^\ell \l^\ell.\] Note the similarity to the
bounds in Theorem~\ref{thm:sum}.  The lower bound in
Theorem~\ref{thm:sum} shows that the upper bound is, in fact, tight.
\end{remark}

\medskip

Now we have

\begin{theorem}\label{log}
Let $H = H(b, \d, \l)$ be a balanced HST, let $W$ be a randomly chosen
$n$-element submultiset of its leaves, and let $h=\min (\d, \log_b
n)$.  Then
\[\E[M(W)] \leq C_{b, \l} \sum_{k=1}^{h} (b\lambda)^k\]
for some positive constant $C_{b, \l}$ depending only on $b$ and $\l$.
\end{theorem}
Note that the upper index of the summation has changed from $\d$
to $h$.

\medskip

\noindent {\bf Proof:} If $\d\le \log_bn$ then there is nothing to prove. 
Hence, we may assume that $\d>\log_b n,$ and consider the case $h = \log_b n$ for $n$ a power of $b$.


Each node at depth $h$ will have an average of one element of $W$ in
its subtree.  The worst case occurs when each node at depth $h$ has
\emph{exactly} one element: otherwise, elements can be paired off in
the subtrees without looking above level $h$.  Thus the situation is
the same as locating points at level $h$ instead of at level $\d$.
More generally, when $n$ is not a power of $b$, the average number of
elements of $W$ in each subtree will still be $O(1)$, giving us the
same result (with a different constant).
The bound then comes from applying Theorem~\ref{thm:sum} to a tree of height $h$.  
\hfill \qedf

\smallskip

Relaxing the index of the second summation in Theorem~\ref{thm:sum} from $\d$
to $h$, we conclude
\begin{theorem}\label{thm:mono_matching}
Let $H = H(b, \d, \l)$ be a balanced HST, let $W$ be a randomly chosen
$n$-element submultiset of its leaves, and let $h=\min (\d, \log_b
n)$.  Then
\[A'_{b, \l} \sum_{k=1}^{h} (b\lambda)^k \leq \E[M(W)] \leq B'_{b, \l} \sum_{k=1}^{h} (b\lambda)^k\]
for some positive constants $A'_{b, \l}, B'_{b, \l}$ depending only on $b$ and $\l$.
\end{theorem}

\bigskip

Thus given an HST with branching factor $b$ and weight ratio $\lambda$,
we can classify the expected matching cost for large $n$ based on the value of $b\lambda$:

\begin{itemize}
\item If $b\lambda < 1$, the expected cost is bounded by a constant.
\item If $b\lambda=1$, the expected cost is $\Theta(h)$.
\item If $b\lambda>1$, the expected cost is $\Theta((b\lambda)^h)$.
When $h = \log_b n$, this is equivalent to $\Theta(n^{1+\log_b \lambda})$.
\end{itemize}

In the proof, we saw that the level $(\log_b n$) of the tree turned out to be crucial
to the tight bound on the matching cost; most of the cost
came from the tree edges above this level, while the edges below had a negligible effect.
This phenomenon will also emerge with the problems we consider later.
We will employ a more general technique, which we call \emph{lifting}, to achieve similar results.

\section{Monochromatic TSP and MST}\label{mono_tsp}

Let us first consider the TSP.  We construct the tour as follows: we
build a tree $T'$ that contains each edge $e = (p, c)$, where $p$ is
the parent vertex of $c$, if and only if some descendant of $c$ is
identified with a point of $W$.  The tour $\cal T$ is formed by
visiting the leaves of $T'$ in order; it is clear that the weight of
$\cal T$ is exactly twice the weight of $T'$, since each edge of $T'$
will be traversed exactly twice to form the tour $\cal T$.

\begin{remark}
It is fairly evident that an optimal tour on $W$ and the root of the HST
can be formed in this way.
Suppose we have a tour that does not have this form; then there are
two leaves $u$ and $v$ that are out of order in the tour.  Switching
the positions of $u$ and $v$ in the tour can only decrease the total
cost.
\end{remark}

Observe that even though the root is not always needed to
create a tour on $W$, it will be needed with high probability.  More
precisely, the probability that it will \emph{not} be needed is the
same as the probability that all $n$ chosen leaves of $T$ have the
same ancestor at level 1 of the tree; this probability is $1/b^{n-1}$,
which is at most $1/2$ for $b, n \geq 2$.
In the following, we presume that the root is needed; this only
changes the true values of the expectations by a constant factor,
which we can ignore.

We now consider the cost of this tour.  Let $v$ be a non-root vertex
in $T$, and let $\ell$ be its level in the tree (with the root being
at level 0).  The probability that a randomly chosen leaf of $T$ is a
descendant of $v$ is $1/b^\ell$.  Since $n$ points will be chosen, the
probability that the parent edge of $v$ is needed in $\cal T$ is
\[1 - \left(1 - \frac{1}{b^\ell}\right)^n.\]
Since there are $b^\ell$ vertices in level $\ell$ and each of their
parent edges has weight $\lambda^\ell$, the total contribution of
level $\ell$ to the weight of $\cal T$ is
\[2 \cdot (b\lambda)^\ell \cdot \left(1 - \left(1 - \frac{1}{b^\ell}\right)^n
\right),\]
hence:
\begin{theorem}
Let $H = H(b, \d, \l)$ be a balanced HST, and let $W$ be a randomly
chosen $n$-element submultiset of its leaves.  Then the expected cost
of a tour on $W$ and the root of $T$ is
\[2 \cdot \sum_{\ell=1}^{\d} \left[(b\lambda)^\ell \cdot \left(1 - 
\left(1 - \frac{1}{b^\ell}\right)^n\right)\right].\]
\end{theorem}

Since removing the root from the tour can only decrease its cost, we
have
\begin{corollary}
\[\E[TSP(W)] \leq 2 \cdot \sum_{\ell=1}^{\d} \left[(b\lambda)^\ell \cdot 
\left(1 - \left(1 - \frac{1}{b^\ell}\right)^n\right)\right].\]
\end{corollary}

\begin{corollary}\label{tsp_upper}
\[\E[TSP(W)] \leq 2 \cdot \sum_{\ell=1}^{\d} (b\lambda)^\ell.\]
\end{corollary}

By the inequality $1 + x \leq e^x$, when $n \geq b^\ell$ we have
\[\left(1 - \frac{1}{b^\ell}\right)^n
\leq e^{-n/b^\ell} \leq \frac{1}{e},\]
hence
\begin{lemma}
Given that the root is necessary in the tour,
\[\E[TSP(W)] \geq 2 \cdot (1 - 1/e) \cdot \sum_{\ell=1}^{h} (b\lambda)^\ell\]
where $h = \min(\d, \log_b n)$.
\end{lemma}

Since the root is necessary with probability at least $1/2$,
\begin{corollary}\label{tsp_lower}
\[\E[TSP(W)] \geq (1 - 1/e) \cdot \sum_{\ell=1}^{h} (b\lambda)^\ell\]
where $h = \min(\d, \log_b n)$.
\end{corollary}

\medskip

Our goal now is to reconcile these two corollaries.
A new notion, which we term ``{lifting}'', will aid us here;
it will also help us later on with the bichromatic problems.
Given an HST $H(b, \d, \l)$ and an $n$-point multiset $W$ with $\d \geq \log_b n$,
we define the {\it lifting of $W$}, written ${\mathcal L}_H(W)$, to be the multiset
formed from the level-$(\log_b n)$ ancestor of each point in $W$.  Stated otherwise,
if we define $A(v)$ to be the ancestor of the point $v$ that lies in level $\log_b n$
of the tree, then \[{\mathcal L}_H(W) = \bigcup_{v \in W} A(v),\]
where each point in level $\log_b n$ is included as many times
as it has descendants in $W$.

For convenience we define $h = \log_b n$.
\begin{claim}
Let $v$ be a vertex of $H$ that sits in level $\ell \ge h.$ Then $$d_H(v, {\mathcal L}_H(v))\le \sum_{i=h+1}^{\infty}\lambda^i=\frac{\lambda^{h+1}}{1-\lambda},$$
where $d_H(x,y)$ denotes the distance of $x$ and $y$ in $H.$
\end{claim}

The claim follows easily from the definition of the edge weights in an HST.  From this claim follows

\begin{lemma}[the lifting lemma]\label{projection}
Assume that a balanced HST $H(b, \d, \l)$ has depth $\d > h.$  Let $W$ be an $n$-element multiset of the leaves.  Then 
$$d_H(W, {\mathcal L}_H(W)) = O((b\lambda)^h).$$ 
\end{lemma}

\noindent{\bf Proof:} $d_H(W, {\mathcal L}_H(W)) \le n \cdot {\lambda^{h+1}}/{(1-\lambda)}=(b\lambda)^h/{(1-\lambda)}.$\hfill\qedf

\medskip

The lifting lemma enables us to use a three-part strategy on these problems when $\d > h$:
\begin{enumerate}
\item Lift the points from the leaves to level $h$.  This incurs a certain cost, say $\alpha$.

\item Solve the problem with the points placed at level $h$.  Let us say the cost is $\beta$.

\item If we can show $\alpha = O(\beta)$, the cost of the original problem is $\Theta(\beta)$.
\end{enumerate}

\medskip


We come now to our main result for the monochromatic TSP.
\begin{theorem}\label{tsp_thm}
\[\E[TSP(W)] = \Theta\left(\sum_{\ell=1}^{h} (b\lambda)^\ell\right)\]
where $h = \min(\d, \log_b n)$.
\end{theorem}
{\bf Proof:}
Assume $h = \log_b n$, since otherwise the theorem follows directly from Corollaries \ref{tsp_upper} and \ref{tsp_lower}.
The lower bound is evident from Corollary \ref{tsp_lower}.
From Lemma~\ref{projection}, the cost of
lifting the points residing at the leaves to level $h$ is $O((b\lambda)^h).$
Since the points are now at level $h$ of the HST, we will treat $h$ as the new height:
by Corollary~\ref{tsp_upper},
the cost of this tour on the lifted points is $\Theta\left(\sum_{\ell=1}^{h} (b\lambda)^\ell\right)$.
The sum of these two costs is $\Theta\left(\sum_{\ell=1}^{h} (b\lambda)^\ell\right)$.

The tour so constructed is not in fact the optimal tour, but instead a ``lazier'' one:
above level $h$ the tour is unchanged, but below level $h$ there are $n$ direct circuits,
one between each point and its ancestor in level $h$.
The cost of the optimal tour is therefore $O\left(\sum_{\ell=1}^{h} (b\lambda)^\ell\right)$.
Since, by Corollary \ref{tsp_lower}, the cost above level $h$ is $\Omega\left(\sum_{\ell=1}^{h} (b\lambda)^\ell\right)$,
the theorem is proved.
\qedf

\bigskip \bigskip \bigskip


Let us turn to the MST problem.  As observed above, with probability
at least $1/2$, the root is needed to form the MST; in what follows we
assume that the root is part of the MST.  Let $F$ be a rooted tree
with positive edge weights, and let $\r$ be its root.  Assume further
that every leaf lies in level $k$ of the tree.  Let us denote by
$d(u,v)$ the total edge weight of the path connecting the leaves $u$
and $v$.  These distances clearly determine a metric space
$\mathcal{M}_F$ on the leaf set of $F.$ For any $x \in V(F),$ let
$F(x)$ denote the subtree of $F$ rooted at $x.$

\begin{lemma}\label{connectedness}
Let $W$ be a submultiset of the leaves of $F.$ Denote by $T$ the
minimum spanning tree on $W$ with distances determined by
$\mathcal{M}_F.$ Also, for an arbitrary non-root vertex $x \in V(F)$, let $T(x)$ be the forest
spanned by the vertices of $V(T) \cap V(F(x)).$ Then the forest $T(x)$ is either connected or empty.
\end{lemma}

\noindent {\bf Proof:} Assume on the contrary that some $T(x)$ is
disconnected and non-empty.
Let $C_1$ and $C_2$ be two of its components.  We will change $T$ in
the following way: delete the edge that connects $C_1$ to $T-T(x),$
and connect $C_1$ to $C_2$ by an edge.  This new tree has a smaller
total edge weight, since we can connect $C_1$ and $C_2$ by keeping the edges of the
$C_1-(T-T(x))$ path from $C_1$ to $x$ and deleting the other edges.
\hfill\qedf

\medskip

We can say more about the structure of the minimum spanning tree if we
impose another condition on $F$:

\begin{lemma}
Assume that all edges that connect a parent with its children have
equal weight.  Then one of the minimum spanning trees is a
path.
\end{lemma}

\noindent {\bf Proof:}
Let us first consider the case when $F$ is a star tree.  In this case,
the distance between any two distinct leaves is the same, so there is an MST
which is a path.

When $F$ is not a star tree, we apply the lemma inductively to the
subtrees rooted at the children of $\r$; each subtree has an MST that
is a path.  Stitching together these MSTs creates an MST of $F$ that
is a path.  To see its minimality, note that if it were not optimal,
the suboptimal pieces could be replaced by optimal ones.  We know that
the MST can be formed piecewise in this way from
Lemma~\ref{connectedness}.  \hfill\qedf


\medskip


The following lemma is immediate.
\begin{lemma}
For any HST $F$, it holds that $0 < TSP(F) - MST(F)
\leq diam(F).$
\end{lemma}
{\bf Proof:} By the preceding lemma, we may take the MST to be a path.
The first inequality follows from the fact that removing an edge from
a Hamiltonian tour yields a spanning tree.  To see the second
inequality, observe that the TSP solution and the MST differ only in
that the former is a cycle and the latter is a chain; the TSP tour is
obtained by adding one edge to the MST.  This edge has weight at most
$diam(F)$.  \hfill\qedf

Notice that $diam(F)= 2\sum_{i=1}^{\delta}\lambda^i < 2\lambda /
(1-\lambda)$, which we consider constant.  The bounds (expectation and, later,
concentration inequalities) obtained for the TSP will therefore
apply to the MST problem as well.

\begin{corollary}
\[0< \E[TSP(W)]- \E[MST(W)] \le diam(F).\]
\end{corollary}

Thus Theorem \ref{tsp_thm} also applies to MSTs, viz.,
\begin{theorem}\label{mst_thm}
\[\E[MST(W)] = \Theta\left(\sum_{\ell=1}^{h} (b\lambda)^\ell\right)\]
where $h = \min(\d, \log_b n)$.
\end{theorem}

\section{Bichromatic MST}\label{bi_mst}

We now turn our attention to the bichromatic problems.
It turns out that the bichromatic MST problem is substantially
different from the bichromatic TSP.  Consider the following problem:
$F$ is a star graph centered at $\r$ and has $l$ leaves, $x_1, x_2,
\ldots, x_l$, with $l$ even.  Assume that at half the leaves $x_i$
there are $s > 1$
red points and one blue point, and at each of the
other leaves there are $s$ blue points and one red point.  We
stipulate that the distance of two points residing at the same leaf is
zero, and the distance between any other pair of points is one.  It is
easy to see that the optimal cost of the bichromatic TSP is $(s-1)l,$
while for a bichromatic spanning tree we will have a total edge length
of $l-1$.

The following observation will be crucial; the reader may find it intuitive, but we state it here to refer
to it later.
\begin{observation}\label{bi-mono}
Let $\mathcal P$ be a monochromatic problem and $\mathcal P'$ be the corresponding bichromatic problem.
Writing $\mathcal{P}(W)$ for the cost of the optimal solution to ${\mathcal P}$ on the point set $W$,
and writing $\mathcal{P'}(R, B)$ for the cost of optimal solution to $\mathcal P'$ on the sets $R$ and $B$,
we have \[\mathcal{P}(R \cup B) \leq \mathcal{P'}(R, B).\]
\end{observation}

Let $R$ and $B$ be two multisets of $n$ points chosen uniformly at
random from the leaves of an HST $H(b, \d, \l)$.  To form the
bichromatic MST, we must connect each red point $r_i$ to its nearest
blue neighbor, and similarly, we must connect each blue point $b_i$ to
its nearest red neighbor.  Since these two cases are symmetric, we
will assume without loss of generality that we are concerned with
connecting a red point $r_i$ to a blue neighbor.

We will write $N(r_i)$ for the closest opposite-colored point in the tree.
(In fact, more than one
  point may be the ``closest''; we can disambiguate by making $N(r_i)$
  the leftmost closest point.)
It is evident that $N(r_i)$ is the blue point $p$ that minimizes the
height of the highest node $h$ in the path from $r_i$ to $p$.  If $p$
is at the same leaf as $r_i$, the cost will be zero.  Otherwise, if
$h$ is in level $\ell$, the cost of the path from $r_i$ to $h$ will be
$\lambda^{\ell+1} + \lambda^{\ell+2} + \ldots + \lambda^\delta <
\lambda^{\ell+1} / (1 - \lambda)$, while the cost of the path from $h$
to $p$ will also be between $\lambda^{\ell+1}$ and
$\lambda^{\ell+1}/(1-\lambda)$.  Thus the cost that $r_i$ contributes
to the bichromatic MST is between $2\lambda^{\ell+1}$ and
$2\lambda^{\ell+1}/(1-\lambda)$.

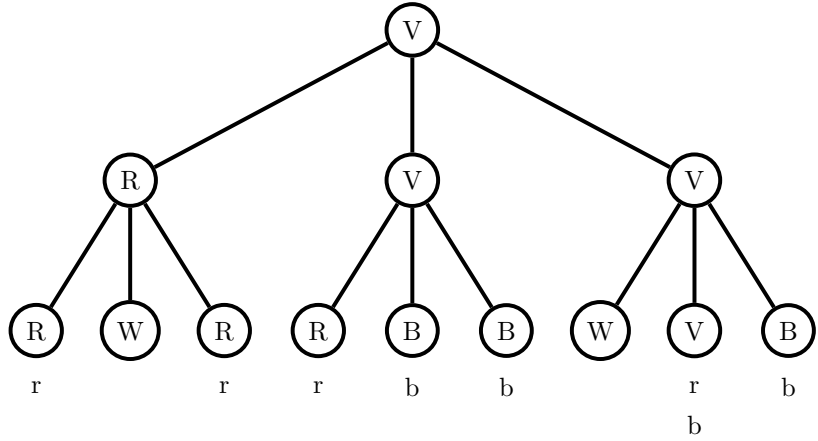
\begin{figure}
\begin{center}
\begin{tikzpicture}[scale=2.5,line width=1.5pt]
    \node (root) at (0, -0.4) [circle,draw] {V};
    
    \node (1) at (-1.5, -1.2) [circle,draw] {R};
    \node (2) at (0, -1.2) [circle,draw] {V};
    \node (3) at (1.5, -1.2) [circle,draw] {V};

    \node (11) at (-2, -2) [circle,draw] {R};
    \node (12) at (-1.5, -2) [circle,draw] {W};
    \node (13) at (-1, -2) [circle,draw] {R};
    \node (21) at (-0.5, -2) [circle,draw] {R};
    \node (22) at (0, -2) [circle,draw] {B};
    \node (23) at (0.5, -2) [circle,draw] {B};
    \node (31) at (1, -2) [circle,draw] {W};
    \node (32) at (1.5, -2) [circle,draw] {V};
    \node (33) at (2, -2) [circle,draw] {B};
    
    \node at (-2, -2.3) {r};
    \node at (-1, -2.3) {r};
    \node at (-0.5, -2.3) {r};
    \node at (0, -2.3) {b};
    \node at (0.5, -2.3) {b};
    \node at (1.5, -2.3) {r};
    \node at (1.5, -2.5) {b};
    \node at (2, -2.3) {b};
    
    \draw (root) -- (1);
    \draw (root) -- (2);
    \draw (root) -- (3);
    \draw (1) -- (11);
    \draw (1) -- (12);
    \draw (1) -- (13);
    \draw (2) -- (21);
    \draw (2) -- (22);
    \draw (2) -- (23);
    \draw (3) -- (31);
    \draw (3) -- (32);
    \draw (3) -- (33);
\end{tikzpicture}
\caption{
Four red points and four blue points are distributed
among the leaves of an HST with $b = 3$ and $\d = 2$.
Lowercase letters represent points; uppercase letters represent the node colors.
R = red, B = blue, V = violet, W = white (blank)
\label{fig:colortree}
}
\end{center}
\end{figure}

We can illustrate this process by coloring the nodes of the MST as
follows.  We color a node violet if it has both red and blue points in
its subtree.  If it has no points in its subtree, we leave the node
white.  Otherwise, the node is colored red or blue, depending on which
color of points appears in its subtree.  See Figure \ref{fig:colortree} for an example (the
colors are represented by the letters R, B, V, W).

The cost of the bichromatic MST is then obtained by finding the cost
from each point $p_i$ to its nearest (ancestral) violet node $V(p_i)$.

There is a caveat with this formulation: the resulting structure may
not be connected.  (This will occur, for example, when each subtree of
the root receives $n/b$ red points and $n/b$ blue points.)  This
difficulty is easily remedied by including the edges of the
\emph{monochromatic} MST on $R \, \cup \, B$ as part of the
bichromatic MST.  The cost of the monochromatic MST does not increase
the asymptotic cost of the bichromatic MST, since the former is easily
seen to cost no more than the latter by Observation~\ref{bi-mono}. 

We can now state
\begin{proposition}\label{initial_bichromatic_mst}
Let $R$ and $B$ be two uniformly chosen multisets of $n$ leaves from
the HST $H(b, \d, \l)$.  The expected cost of the bichromatic MST on
$R$ and $B$ is
\[\Theta(\mu + n \cdot \E[{\cal C}(b, \d, \l, n)]),\]
where ${\cal C}(b, \d, \l, n)$ is the cost of the path from a point
$p$ to its lowest violet ancestor in $H$, and $\mu$ is the expected
cost of the monochromatic MST of $R \cup B$.
\end{proposition}

Now we turn to the task of estimating $\E[{\cal C}(b, \d, \l, n)]$.
Let the first red point $r_1$ be located, without loss of generality,
at the leftmost leaf $L$ of the HST.  We define the function $P(m, n,
i, j)$ to be the probability that the smallest element of an
$n$-element randomly chosen submultiset of $\{1, \ldots, m\}$ falls in
the range $i, \ldots, j$.  Then

\begin{itemize}
\item the probability that its leaf $L$ is violet is the probability
  that one of the blue points is also located at $L$, which is
  $P(b^\d, n, 1, 1) = 1 - (1-1/b^\d)^n$.

\item The probability that $L$ is not violet, but its parent \emph{is}
  violet, is the probability that $L$ has no blue point but one of its
  siblings does.  This probability is $P(b^\d, n, 2, b)$.

\item The probability that $L$ and its parent are not violet, but its
  grandparent \emph{is} violet, is $P(b^\d, n, b+1, b^2)$, etc., etc.
\end{itemize}


\begin{lemma} \[P(m, n, i, j) = \frac{(m-i+1)^n -
  (m-j)^n}{m^n}.\]
\end{lemma}

\noindent {\bf Proof:} One can visualize the elements of the multiset as lattice points in the
$n$-dimensional cube with edge length $m.$ Then the total number of points is $m^n.$
The chosen points must belong to a sub-cube with edge length $m-i+1$; however, we have to discard
those lattice points that have only large entries. These lattice points belong to a sub-cube of
volume $(m-j)^n.$ 
\hfill\qedf

\medskip

In general, the probability that the red point will have its lowest
violet ancestor in the $\ell$-th level from the \emph{bottom} of the
tree, $1 \leq \ell \leq \d$, is
\begin{eqnarray*}
P(b^\d, n, b^{\ell-1} + 1, b^{\ell}) &=& \frac{(b^\d - b^{\ell-1})^n -
  (b^\d - b^{\ell})^n}{(b^\d)^n} \\ &=& (1 - b^{(\ell - \d) - 1})^n - (1 - b^{\ell-\d})^n.
\end{eqnarray*}

Clearly $P(b^\d, n, b^{\ell-1} + 1, b^{\ell}) \leq (1 - b^{(\ell - \d) - 1})^n$.
Under stronger conditions, a similar lower bound holds (Lemma \ref{P_bound} below).
Before we prove it, we must make a brief digression to talk about the behavior of
the expression $(1 \pm r/n)^n$ for $r$ fixed.  It is well known that it approaches $e^{\pm r}$ as $n \to \infty$.
But since our earlier results did not rely on $n$ going to infinity, we would like to avoid that here as well.
Fortunately, the expression is bounded between exponentials even for relatively small values of $n$.
\begin{proposition}\label{e_bound}
Let $r > 0$.  Then for $n > r$, \[e^{r/2} < (1+r/n)^n < e^r,\]
and for $n > 3r$, \[e^{-3r/2} < (1-r/n)^n < e^{-r}.\]
\end{proposition}
{\bf Proof:}
The upper bounds follow from the elementary inequality $1 + y < e^y$ for $y \neq 0$.
(The simplest proof: The expression $e^y - (1+y)$ has its unique global minimum at $y=0$, shown
by taking derivatives.  The value of this minimum is 0.)

Now we show the lower bounds.  We use the Taylor series \[\ln (1-x) = -x-x^2/2-x^3/3-\ldots,\]
valid for $-1 \leq x < 1$.
\begin{eqnarray*}
n \cdot \ln(1+r/n) &=& n \cdot (r/n - r^2/2n^2 + r^2/3n^3 - r^4/4n^4 + \ldots) \\
&>& n \cdot (r/n - r^2/2n^2) \\
&=& r - r^2/2n.
\end{eqnarray*}
Simplifying $n - r^2/n > r/2$ shows that it is equivalent to $n > r$.  Thus for $n > r$, we have
$n \cdot \ln(1+r/n) > r/2$.  Exponentiating both sides gives the first statement.

Similarly,
\begin{eqnarray*}
n \cdot \ln(1-r/n) &=& n \cdot (-r/n-r^2/2n^2-r^3/3n^3-\ldots) \\
&>& n \cdot (-r/n-r^2/n^2-r^3/n^3-\ldots) \\
&=& n \cdot \frac{-r/n}{1-r/n} \\
&=& \frac{rn}{r-n}.
\end{eqnarray*}
Simplifying $rn/(r-n) > -3r/2$ shows that it is equivalent to $n > 3r$.  Thus for $n > 3r$, we have
$n \cdot \ln(1-r/n) > -3r/2$.  Exponentiating both sides gives the second statement.
\qedf

\medskip

We now return to the lower bound on $P(b^\d, n, b^{\ell-1} + 1, b^{\ell})$.
\begin{lemma}\label{P_bound}  When $\d \leq \log_b n$ and $n > 1$,
\[P(b^\d, n, b^{\ell-1} + 1, b^{\ell}) \geq 1/5 \cdot (1 - b^{(\ell - \d) - 1})^n.\]
\end{lemma}
{\bf Proof:}
For brevity, we write $z = b^{\ell-\d}$.  
The smallest that $z$ can be is $b^{-\d} \geq b^{-\log_b n} = 1/n$ --- that is, $z$ cannot be smaller than $1/n$.

Consider the fraction $\displaystyle\frac{1-z}{1-z/b}$.  It is less than 1, and its value increases as $z$ decreases.
Thus \[\displaystyle\left(\frac{1-z}{1-z/b}\right)^n \leq \frac{(1-1/n)^n}{(1-1/bn)^n}
\leq \frac{e^{-1}}{e^{-3/2b}} = e^{3/2b-1} \leq e^{-1/4} < 4/5.\]
(We have used Proposition \ref{e_bound} to show $(1-1/bn)^n \geq e^{-3/2b}$.  Notice that $n > 1$ implies $n > 3/b$.)

So
\[(1-z)^n \leq 4/5 \cdot (1-z/b)^n,\] and \[(1-z/b)^n - (1-z)^n \geq 1/5 \cdot (1-z/b)^n.\]
\qedf

Accordingly, given $\d \leq \log_b n$, we have

\begin{eqnarray*}
\E[{\cal C}(b, \d, \l, n)] &=& \Theta\Big(0 \cdot P(b^\d, n, 1, 1) +
2\lambda^\d / (1 - \l) \cdot P(b^\d, n, 2, b) + \\ && 2\lambda^{\d-1}
/ (1 - \l) \cdot P(b^\d, b, b+1, b^2) + 2\lambda^{\d-2} / (1 - \l)
\cdot P(b^\d, n, b^2+1, b^3) + \ldots + \\ && 2\lambda^1 / (1 - \l)
\cdot P(b^\d, n, b^{\d-1} + 1, b^\d)\Big) \\
&=& \Theta\left(\sum_{\ell=1}^\d 2\l^{(\d
  - \ell) + 1} / (1 - \l) \cdot (1 - b^{(\ell - \d) - 1})^n\right) \\
&=& 2\l^\d \cdot \Theta\left(\sum_{\ell=1}^\d \l^{1 - \ell}
\cdot (1 - b^{(\ell - \d) - 1})^n\right)
\end{eqnarray*}
for fixed $b$ and $\l$.
Our next task is to show
that the expression $\Theta\left(\sum_{\ell=1}^\d \l^{1 - \ell} \cdot (1 - b^{(\ell - \d) - 1})^n\right)$ that appears above
is bounded between constants.
\begin{lemma}\label{bela} 
Let $b \geq 2$ and $\lambda < 1$ be fixed, let $\d \leq \log_b n$, and let $n > 4 \ln (1/\l)$.
Define $a_\ell = \l^{1 - \ell} \cdot (1 - b^{(\ell - \d) - 1})^n$.
Then there is a constant $C$ such that \[\frac{a_\ell}{a_{\ell+1}} \geq \frac 1 \l\]
for all $\ell \in \{C, \ldots, \d\}$.
\end{lemma}
{\bf Proof:}
Defining $z = b^{\ell - \d}$ as above, we have
\begin{eqnarray*}
\frac{a_\ell}{a_{\ell+1}} &=& \lambda \left(\frac{1-z/b}{1-z}\right)^n \\
&\geq& \l \left(\frac{1-z/2}{1-z}\right)^n \\
&=& \l \left(1 + \frac{z/2}{1-z}\right)^n \\
&\geq& \l \left(1 + \frac z 2\right)^n.
\end{eqnarray*}
When $z > (8 \ln (1/\l))/n$, we have
\[\left(1 + \frac z 2\right)^n > \left(1 + \frac{4\ln(1/\l)}{n}\right)^n
> e^{2\ln(1/\l)} = \frac{1}{\l^2};\]
thus $\left(1 + \frac z 2\right)^n \geq \frac{1}{\l^2}$,
and we have $a_\ell/a_{\ell+1} \geq 1/\l$.

We have shown that $a_\ell/a_{\ell+1} \geq 1/\l$ can fail to hold only when $z \leq (8 \ln (1/\l))/n$.
We need to see what this means in terms of $\ell$:
\begin{eqnarray*}
z &\leq& \frac{8\ln(1/\l)}n \\
b^{\ell-\d} &\leq& \frac{8\ln(1/\l)}n \\
\ell-\d &\leq& \log_b \frac{8\ln(1/\l)}n \\
\ell &\leq& \log_b \frac{8\ln(1/\l)}n + \d \\
&\leq& \log_b \frac{8\ln(1/\l)}n + \log_b n \\
&=& \log_b (8\ln(1/\l)).
\end{eqnarray*}
Thus $a_\ell/a_{\ell+1} \geq 1/\l$ is true with the possible exceptions of when $\ell = 1, 2, \ldots, \log_b (8\ln(1/\l))$.
Set $C = \log_b (8\ln(1/\l)) + 1$.
\qedf

We next consider the behavior of the expression $n \cdot (1-1/b^\d)^n$.  It will be more convenient to substitute
$z = b^\d$.
\begin{lemma}\label{tom_ineq}
Let $z \geq 1$ and $n > 0$.  Then \[n \cdot (1 - 1/z)^n < z.\]
\end{lemma}
{\bf Proof:}
If $z=1$ the proof is trivial, so let us assume $z > 1$.
Dividing and taking roots, we obtain the equivalent statement $1-1/z < \sqrt[n]{z/n}$.
We will prove \[1-\frac 1 z \leq e^{-1/z} < e^{-1/ez} \leq \sqrt[n]{\frac z n}.\]
The first inequality of these three follows from the well-known fact $1+y \leq e^y$, here taking $y = -1/z$.
The second is equally simple, since $a < b$ implies $e^a < e^b$ (here we take $a = -1/z$ and $b = -1/ez$).

Only the third inequality remains to be proved.
We will show that the minimum value of $\sqrt[n]{z/n} = (z/n)^{1/n}$ as a function of $n$ (considering $z$ constant)
is $e^{-1/ez}$.
Several manipulations will make this task easier.  First, set $x = 1/n$.  Then \[(z/n)^{1/n} = (zx)^x,\]
where $x \in (0, \infty)$.  It is equivalent to minimize its logarithm, \[\ln (zx)^x = x \ln zx = x \ln x + x \ln z.\]
This expression has derivative \[1 + \ln x + \ln z;\] the derivative is zero when $x = 1/ez$.
This value of $x$ does, in fact, give a minimum, since the second derivative $1/x$ is positive.
We conclude that the minimum value of $(zx)^x$ is \[(z/ez)^{1/ez} = e^{-1/ez},\]
as desired.
\qedf

\begin{corollary}
Let $b \geq 2$ and $\lambda < 1$ be fixed, and let $\d \leq \log_b n$.  Then
\[n \cdot \sum_{\ell=1}^\d \l^{1 - \ell} \cdot (1 - b^{(\ell - \d) - 1})^n = O(b^\d).\]
\end{corollary}
{\bf Proof:}
Let $n$ be sufficiently large.
We define $a_\ell = \l^{1 - \ell} \cdot (1 - b^{(\ell - \d) - 1})^n$.
Notice $a_1 = (1 - b^{-\d})^n$.  By Lemma \ref{tom_ineq} with $z = b^\d$,
we have $n \cdot a_1 < b^\d$.  For $\ell > 1$, we have $(1 - b^{(\ell - \d) - 1})^n < (1 - b^{-\d})^n$,
so that $n \cdot a_\ell < \l^{1-\ell} \cdot b^\d$.

Set $C$ as in Lemma \ref{bela}.
We see that each of the first $C$ terms $n \cdot a_1, \ldots, n \cdot a_C$
is at most $\l^{-C} \cdot b^{\d}$, so their sum is $\leq C\l^{-C} b^{\d}$.
For $\ell > C$, we have $a_{\ell+1}/a_\ell \leq \lambda$, so that
the sum $n \cdot \sum_{\ell=C+1}^\d a_\ell$ converges faster than a geometric series with common ratio $\lambda$;
this part of the sum is $\leq C\l^{-C}b^{\d}/(1-\l)$.  Both these quantities are $O(b^{\d})$.
\qedf

We can now state
\begin{proposition}
For $\d \leq \log_b n$, \[\E[n \cdot {\cal C}(b, \d, \l, n)] = O((b\lambda)^\delta).\]
\end{proposition}
{\bf Proof:}
\begin{eqnarray*}
\E[n \cdot {\cal C}(b, \d, \l, n)] &=& n \cdot 2\l^\d \cdot \Theta\left(\sum_{\ell=1}^\d \l^{1 - \ell}
\cdot (1 - b^{(\ell - \d) - 1})^n\right) \\
&=& n \cdot 2\l^\d \cdot O(b^\d / n) \text{ \,\,\,\,\,\,\,\,\,\,\, (by the previous Corollary)}\\
&=& O(\l^\d b^\d).
\end{eqnarray*}
\qedf

Applying this result to Proposition~\ref{initial_bichromatic_mst}
gives our main result for the bichromatic MST.
\begin{theorem}\label{bichrom_mst}
Let $R$ and $B$ be two uniformly chosen multisets of $n$ leaves from
the HST $H(b, \d, \l)$.
The expected cost of the bichromatic MST on
$R$ and $B$ is
\[\Theta\left(\sum_{i=1}^h (b\lambda)^i \right),\]
where $h = \min (\d, \log_b n)$.
\end{theorem}
This is the same (up to constant factors depending on $b$ and $\l$)
as the expected cost for the monochromatic MST on $R \cup B$.

\medskip

\noindent {\bf Proof:}
Suppose $\d \leq \log_b n$.  
By Proposition \ref{initial_bichromatic_mst}, the expected cost
is $\Theta\left(\E[n \cdot {\cal C}(b, \d, \l, n)] + \sum_{i=1}^\d (b\lambda)^i \right)$;
by Observation \ref{bi-mono}, it is $\Omega\left(\sum_{i=1}^\d (b\lambda)^i \right)$.
We need only to show $\E[n \cdot {\cal C}(b, \d, \l, n)] = O\left(\sum_{i=1}^\d (b\lambda)^i \right)$.
But by the previous Proposition, $\E[n \cdot {\cal C}(b, \d, \l, n)] = O((b\lambda)^\delta)$,
and the last term in the summation is $(b\lambda)^\delta$.  Thus
the total cost is also $O\left(\sum_{i=1}^\d (b\lambda)^i \right)$.

If $\d > \log_b n$, we lift the points up to level $\log_b n$.
By the above argument,
the expected cost of the bichromatic MST on the lifted points
is $\Theta\left(\sum_{i=1}^{\log_b n} (b\lambda)^i \right)$.
The lifting cost is $\Theta((b\lambda)^{\log_b n})$, and as before,
such a term is already present in the summation.\qedf

%



\section{Bichromatic TSP}\label{bi_tsp}
 
As before, we start by considering the case when $F$ is a star with edge weight
 $\lambda.$  Let $n$ red points and $n$ blue points be uniformly distributed
 at random among the leaves of $F.$
 We will call a leaf \emph{easy} if it has been assigned equal numbers of red and blue points;
 otherwise, we will call the leaf \emph{hard}.
 Note that a hard leaf can be either monochromatic or bichromatic.
 
 Our optimal traveling salesman tour is produced by the
 following algorithm:

 \begin{itemize}

 \item[1.] Connect the points at every easy leaf into a path that
   begins at a red point and ends at a blue point. These paths have
   zero cost.

 \item[2.] Connect the paths at the easy leaves into one long red-blue
   path. If there are $k$ easy leaves, the total cost of this is
   $(k-1)\lambda.$

 \item[3.] At each hard leaf, find the longest possible red-blue path,
 depending on whether we have more red or more blue points.
 
Since one color outnumbers the other, each of these paths can have the same color at both its endpoints.
   We delete the inner points of all such paths, leaving only the
   two same-colored endpoints.  Hence, after this process, some points are isolated,
   while the other points are all arranged into a long path.

 \item[4.] Connect the isolated points into a long red-blue path, and
   then glue the two long paths together to get the traveling salesman tour.

 \end{itemize}

 As we found with the bichromatic MST, this algorithm can be easily extended
 to find the optimal TSP tour for general HSTs (not just star trees).
 To this end, it is useful to consider a version that
 works when the numbers of red and of blue points are different. In
 such a case, we can only have a path of minimum cost, hence, in Steps 3
 and 4 one has to modify the method to first connect as many
 isolated red-blue points together into a red-blue path as possible,
 and then connect this long path to the path containing the points of
 the easy leaves. 
 
 Notice that in Step 3, the cost of connecting the isolated points
 into a red-blue path is twice the cost of the bichromatic matching
 on these points minus $\lambda.$  Clearly, the discrepancy of the
 point distribution plays an important role here, just as it did in the case of
 bichromatic matching in \cite{Abrahamson}.  
 
 We say that the random variable $Y$ has distribution $Binomial(n,p)$ if $Y=Y_1+Y_2+\ldots+Y_n,$ where
 $P(Y_i=1)=p$ and $P(Y_i=0)=1-p$ for every $1\le i\le n$, with the $Y_i$ independent.
 For a vertex $v$ in level $\ell$,
 the discrepancy has the form $|R(v) - B(v)|$,
 where $R(v)$ and $B(v)$, the numbers of red and blue points in $v$'s subtree, are independent variables
 with distribution $Binomial(n, 1/b^\ell)$.  We now consider the expected behavior of this discrepancy.
 We prove two bounds: one will be useful in the case of a star tree, while the other
 will be used in the bound for general HSTs.

\begin{lemma}\label{abs_bound}
Let $X$ be the difference of two i.i.d.~$Binomial(n, p)$ variables.  For fixed $p$,
\[\E|X| = \Theta(\sqrt n).\]
\end{lemma}

\begin{lemma}\label{abs_bound_2}
Let $X$ be the difference of two i.i.d.~$Binomial(n, p)$ variables, with $n \geq 1/p$ and $p \leq 1/2$.  Then
\[\sqrt {Cnp} \leq \E|X| \leq \sqrt {2np}\] for some constant $C > 0$.
\end{lemma}

\noindent {\bf Proof:}
We denote by $\D Y$ the standard deviation of the random variable $Y$.
From \[0 \leq \Var|X| = \E|X|^2 - (\E |X|)^2 = \E X^2 - (\E |X|)^2 = (\D X)^2 - (\E |X|)^2,\]
we obtain $\E |X| \leq \D X$.  Here we used the fact $\E X=0$, since it is the difference of two random variables with the same distribution. 
Since $\D X = \sqrt{2np(1-p)}$, the upper bounds on $\E |X|$ follow.

The lower bound will take more work.
Recall H\"{o}lder's inequality,
\[\E|YZ| \leq (\E|Y|^q)^{1/q} \cdot (\E|Z|^r)^{1/r} {\rm \,\, when \,\,} 1/q + 1/r = 1.\]
Setting $Y = |X|^{2/3}, Z = |X|^{4/3}, q = 3/2, r = 3$,
we obtain
$\E{|X|^2} \leq (\E |X|)^{2/3}(\E |X|^4)^{1/3}$, and thus
\[\E |X| \geq \frac{(\E X^2)^{3/2}}{\sqrt {\E X^4}}.\]
By definition, we can write $X$ as the sum $X_1 + X_2 + \ldots + X_n$, where the $X_i$'s are i.i.d.~with
\[X_i = \begin{cases}
\hspace{0.3cm}1 & {\rm with \,\, probability \,\,} p(1-p), \\
-1 & {\rm with \,\, probability \,\,} p(1-p), \\
\hspace{0.3cm}0 & {\rm with \,\, probability \,\,} p^2+(1-p)^2.
\end{cases}\]
It is clear that $\E X^2 = 2np(1-p)$.

Consider the expansion
$X^4 = (X_1 + X_2 + \ldots + X_n)^4$.
Any term with an odd power of an $X_i$, such as $X_1^3 X_2$ or $X_1^2 X_2 X_3$,
will have zero expectation, by independence and the fact $\E X_i = \E X_i^3 = 0$. 
Thus the only terms with nonzero expectation
will be the terms 
$X_i^4$ and $X_i^2 X_j^2$ with $i \neq j$.
Since $\E X_i^4 = \E X_i^2= 2p(1-p)$ and $\E X_i^2 X_j^2 = \E X_i^2 \cdot \E X_j^2 = 4p^2(1-p)^2$ for $i \neq j$, we have
\[
\E X^4 = \sum_{i=1}^n X_i^4 + 6\sum_{i \neq j} \E X_i^2 X_j^2
= 2np(1-p) + 24n(n-1)p^2(1-p)^2 = \Theta(n^2).\]

Thus \[\E |X| \geq \frac{(\E X^2)^{3/2}}{\sqrt {\E X^4}} = \frac{[2np(1-p)]^{3/2}}{\sqrt{\Theta(n^2)}}
= \Omega(\sqrt n),\]
proving Lemma \ref{abs_bound}.
%
%
Note that without the $2np(1-p)$ term in the denominator, this function would behave as
$\sqrt{Cnp(1-p)}$.  To see when the $24n(n-1)p^2(1-p)^2$ term will dominate, we set
\[np(1-p) \leq 24n(n-1)p^2(1-p)^2\] and obtain \[1 \leq 24(n-1)p(1-p).\]
For $n \geq 2$ we have $2(n-1) \geq n$, and for $p \leq 1/2$ (as it always is for us)
we have $p(1-p) \geq p/2$; thus $1 \leq np$ will imply $1 \leq 4(n-1)p(1-p)$.
When $n \geq 1/p$, then, this bound is at least $\sqrt{Cnp}$, and Lemma \ref{abs_bound_2}
is proved.
\qedf

\begin{remark}\label{np_remark}
For a vertex in level $\ell$ of the tree, $p = 1 / b^\ell$; thus when $\ell \leq \log_b n$, we have
$n \geq 1/p$.
\end{remark}

Let us now apply Lemma \ref{abs_bound} to the problem of finding the expected TSP cost
for a star tree.
Let the number of red (resp. blue) points assigned to leaf $l_i$ be $R_i$ (resp. $B_i$).
Then $X$ in Lemma~\ref{abs_bound} will be $R_i-B_i$ for the leaf $l_i.$
As described initially, we make as long a path as possible in each leaf;
these paths have zero cost.  Our task is now to connect these paths into a tour.

If $R_i \neq B_i$, then our tour must visit the leaf exactly $|R_i - B_i| + 1$ times.
If $R_i = B_i$, then the cost depends on whether their value is zero or not: if it is zero, then there
is no cost, while if the value is nonzero, the leaf must be visited once.
Thus the cost contributed by leaf $l_i$ is either $|R_i - B_i|$ or $|R_i - B_i| + 1$.

The extra $+1$ does not matter much; by Observation~\ref{bi-mono}, the cost of the optimal \emph{monochromatic} tour
is less than or equal to the cost of the optimal bichromatic tour.  Thus we can add the cost of the
monochromatic tour to the bichromatic tour and only increase its cost by a constant factor.
Since $\E |R_i - B_i| = \Theta(\sqrt n)$ by Lemma~\ref{abs_bound}, we obtain
\begin{proposition}
Let $R$ and $B$ be two multisets of size $n$ of the leaves of the star tree $F = (b, 1, 1)$.
Consider $b$ fixed, and assume that $n$ is large.
The expected cost of the optimal bichromatic tour on $R$ and $B$ is $\Theta(\sqrt n).$
\end{proposition}


The case of a general HST can be handled similarly, this time using Lemma \ref{abs_bound_2}.
Let $v$ be a non-leaf vertex in the HST.  
As described above, we make as long a path as possible in the subtree rooted at $v$.  
The endpoints of this path, and the leftover points, must now look upward to find neighbors in the tour.

Let the number of red (resp. blue) points assigned to a non-leaf node $v$ be $R(v)$ (resp. $B(v)$).
If $R(v) \neq B(v)$, our tour must use $v$ exactly $|R(v) - B(v)| + 1$ times.
If $R(v) = B(v)$, then the cost depends on whether their value is zero or not: if it is zero, then there
is no cost, while if the value is nonzero, the leaf must be used once.
Thus the cost contributed by $v$ is either $|R(v) - B(v)|$ or $|R(v) - B(v)| + 1$.

As before, the extra $+1$ does not matter: using Observation~\ref{bi-mono}, we can add the cost of the
monochromatic tour to the bichromatic tour and only increase its cost by a constant factor.

\begin{theorem}\label{bichrom_tsp}
Let $R$ and $B$ be two multisets of size $n$ of the leaves of the HST $H(b, \d, \l)$.
Consider $b$ and $\l$ fixed, and define $h = \min(\d, \log_b n)$.
Then the expected cost of the optimal bichromatic tour on $R$ and $B$ is 
\[\Theta\left( \sqrt{n}\cdot \sum_{i=1}^h(\sqrt{b}\lambda)^i\right).\]
\end{theorem}



\noindent {\bf Proof:}
Let $v$ be a node in level $\ell \leq h$.  By Lemma \ref{abs_bound_2} and Remark \ref{np_remark},
the expected contribution of $v$ to the cost of the tour is $\Theta(\l^\ell \cdot \sqrt {n/b^\ell}) = \Theta((\l/\sqrt b)^\ell \cdot \sqrt n)$.
The total cost of level $\ell$, obtained by multiplying by $b^\ell$, is therefore $\Theta((\sqrt b \l)^\ell \sqrt n)$;
summing up over levels $1, 2, \ldots, h$ and adding in the cost of the monochromatic MST gives us bounds
of \[\Omega\left(\sqrt{n}\cdot \sum_{i=1}^h(\sqrt{b}\lambda)^i\right) \text{ and } O\left(\sqrt{n}\cdot \sum_{i=1}^h(\sqrt{b}\lambda)^i + \sum_{i=1}^h (b\l)^i\right)\]
for the portion of the tour above level $h$.  But in fact \[\sum_{i=1}^h (b\l)^i = O\left(\sqrt{n}\cdot \sum_{i=1}^h(\sqrt{b}\lambda)^i\right),\]
because $b^i \leq \sqrt{n \cdot b^i}$ for $i \leq h$.  (The inequality simplifies to $b^i \leq n$; recall $b^h = n$.)
We have shown, then, that the expected cost of the tour above level $h$ is \[\Theta\left(\sqrt{n}\cdot \sum_{i=1}^h(\sqrt{b}\lambda)^i\right).\]

If $h = \d$ we are done.  If instead $h = \log_b n$, we must first lift the points to level $h$; the cost is $O((b\lambda)^h)$,
but as above, $(b\lambda)^h = O\left(\sqrt{n}\cdot \sum_{i=1}^h(\sqrt{b}\lambda)^i\right)$.
\hfill\qedf

\bigskip \bigskip

We conclude this section by remarking that our results can also be applied to the corresponding classical probabilistic
optimization problems on the $d$-dimensional Euclidean hypercube.
This is accomplished by using HSTs that dominate the 
distances in a sufficiently dense grid of the cube. It is easy to see that this process
produces upper bounds for the expected costs which are tight up to a constant
factor, except in the case of the 2-dimensional bichromatic matching problem.
For more details on the dominating HSTs, see \cite{Abrahamson}.

\section{Concentration inequalities}\label{concentration}


Many of the ideas in this section stem from those of
Yukich~\cite[Chapter 6]{Yukich}, who in turn used methods of
Talagrand~\cite{Talagrand95}, Rhee~\cite{Rhee93} and
Steele~\cite{Steel97}.  Given a hierarchically separated tree $T$ with
branching factor $b$ and weight ratio $\lambda$ satisfying $0<\lambda <
1$ and $b \lambda \ge 1$, we investigate the probability that the matching length of a
random point set $X$ deviates widely from its expectation.
We will consider both the
monochromatic case and the bichromatic case.  As it turns out, we have
much better concentration results for the former than for the latter,
since in the monochromatic case we can use an isoperimetric inequality
which, it seems, cannot be applied to the bichromatic problems.

First we briefly discuss Azuma's inequality, which will be of use to
us in both cases.  Let $(\Omega, {\mathcal A}, P)$ be a finite
probability space with the filtration (in this case a sequence of
partitions of $\Omega$)
$$(\emptyset, \Omega) = A_0 \subset {\mathcal A}_1 \subset \ldots
\subset {\mathcal A}_t = {\mathcal A}.$$Let $X$ be a random
variable.  For each $1 \le i \le t$ we define the {\it martingale
  difference} $d_i={\E}(X|{\mathcal A}_i)-{
  \E}(X|{\mathcal A}_{i-1}),$ and assume that $\|d_i\|_{\infty}\le
\sigma_i.$ We have the following well-known result:

\begin{theorem}[Azuma's inequality]\label{azuma}
For all $a>0$,
$$P(|X-{\E}X|\ge a)\le 2e^{-a^2 / 2\sigma^2},$$
where $\sigma^2 \equiv \sum_{i=1}^t \sigma^2_i.$ 
\end{theorem}

\subsection{The monochromatic case} 

In this section we investigate the monochromatic case.  First we will consider the minimum matching problem, 
then the closely related TSP and MST problems. 

%

\medskip

We need a combinatorial lemma. 
\begin{lemma}\label{Top}
Assume that $k$ points are assigned to vertices of $T,$ with $k$
even. Then the minimum matching for this point set has total cost at
most $2\cdot Top(k) / (1-\lambda),$ where $Top(k)$ is the sum of the edge
lengths of the first $\lceil \log_b k \rceil$ levels of $T.$
\end{lemma}

\noindent{\bf Proof:} Let $y_1$ and $y_2$ be two points below level $\ell=\lceil \log_b k \rceil$ that are matched in the minimum 
matching. Denote their ancestors at level $\ell$ by $x_1$ and $x_2,$ respectively. Then $d_T(y_1, y_2) \le d_T(x_1, x_2)+2\lambda^{\ell}/(1-\lambda)$ by
Lemma~\ref{projection}. If one of the points, say $y_1,$ is at level $\ell,$ we still have the same inequality.
It is easy to see that no edge of the tree is used more than once in a minimum matching.
Hence, the minimum matching length is at most $Top(k)+k \lambda^{\ell}/(1-\lambda),$ since we have $k/2$ pairs to be matched.

It is easy to see that $Top(k)=b\lambda+\ldots b^{\ell}\lambda^{\ell} \ge k \lambda^{\ell}.$ This implies that the minimum matching length
is at most $2\cdot Top(k)/(1-\lambda).$ 
\hfill \qedf

\medskip

We are now going to use isoperimetric methods in order to prove strong
concentration inequalities.  Toward this end, let $(\Omega, {\mathcal
A}, \mu)$ be the finite probability space with the atoms of $\Omega$
corresponding to the leaves of $T$, with each atom having equal probability.
We define $\Omega^n$ to be the
$n$-fold product space on $\Omega$, and we denote by $\mu^n$ its probability measure.

Given $X, Y \in \Omega,$ the Hamming distance $H$ between $X$ and $Y$
is the number of coordinates in which $X$ and $Y$ disagree:
$$H(X, Y) = |\{i: X_i \neq Y_i\}|.$$

With the following lemma we show that if two $n$-tuples are close in
Hamming distance, the corresponding minimum matching costs are
close to each other. This property is called the {\it smoothness} of
the minimum matching functional. 

\begin{lemma}\label{smoothness}
Let $X, Y \in \Omega^n.$ If $H(X, Y)=k$ then $$M(X) \le M(Y) + 2\cdot
Top(k) / (1-\lambda),$$ where $Top$ is as defined in Lemma~\ref{Top}.
\end{lemma}

\noindent {\bf Proof:} Assume that we have a minimum matching for $Y,$
and then delete/add $k$ points in order to make $X.$ We construct a
matching (not necessarily minimum) for $X$ that will satisfy the
inequality of the lemma.

There are three kinds of matched pairs in $Y.$  First, there are those
that are unaffected --- that is, both points belong to $X$ as well. We
will keep them matched in the new matching.  Second, there are matched
pairs from which we deleted one point each. The remaining points of these pairs became unmatched,
as well as the points of $X-Y.$ Since $H(X, Y)=k,$ we have $k$ unmatched
points (here $k$ must be even). We find a minimum matching
for these points; the cost is at most $2 \cdot Top(k) /
(1-\lambda)$ by Lemma~\ref{Top}. In this way, we
constructed a matching for $X$ having total length at most $M(Y) +
2\cdot Top(k) / (1-\lambda),$ and the lemma is proved. \hfill\qedf

\medskip

Let us consider the smoothness of the TSP functional on HSTs. Earlier we saw that $|TSP-MST|< 2\l/(1-\l),$
which is a constant. Hence, the concentration we show below for the TSP holds for the MST as well.
First we need an analogue of Lemma~\ref{Top}.
\begin{lemma}\label{Top-TSP}
Assume that $k$ points are assigned to vertices of $T,$ with $k$
even. Then the traveling salesman tour for this point set has total cost at
most $4\cdot Top(k) / (1-\lambda),$ where $Top(k)$ is the sum of edge
lengths of the first $\lceil \log_b k \rceil$ levels of $T.$
\end{lemma}

\noindent{\bf Proof:} The proof is very similar to the proof of Lemma~\ref{Top}, so we emphasize the differences only.
First, the TSP tour that connects all the vertices in level $k$ has total length $2\cdot Top(k).$ We use the projection lemma again to get that
for points below level $k$ we have to add at most an extra cost of $2\l/(1-\l).$ Since there are $k$ points, we can have at most $2k\l/(1-\l)$ extra
cost. This adds up to $2\cdot Top(k)+2k\l/(1-\l)\le 4 \cdot Top(k)/(1-\l).$  
\hfill \qedf

\smallskip

With this we are prepared to show the smoothness of the TSP.

\begin{lemma}\label{smoothness-TSP}
Let $X, Y \in \Omega^n.$ If $H(X, Y)=k$ then $$TSP(X) \le TSP(Y) + 8\cdot Top(k) / (1-\lambda).$$ 
\end{lemma}

\noindent {\bf Proof:} Assume that we have an optimal tour $\mathcal T$ for $Y,$
and then delete/add $k$ points in order to make $X.$ We construct a
tour (not necessarily optimal) for $X$ that will satisfy the
inequality of the lemma.

Notice that even if we delete points from ${\mathcal T}$, the optimal cost of the tour ${\mathcal T}_1$
that skips the deleted points, going through only the remaining points in their original order, is at most as large as the cost of $\mathcal T.$
This follows from the triangle inequality: say, $u, v, w$ are in this order in the optimal tour, and then we delete $v.$ Then $d(u,v)+d(v,w)\ge d(u,w).$

Next, we construct an optimal tour ${\mathcal T}_2$ through the new points, with cost at most $4\cdot Top(k)$ by Lemma~\ref{Top-TSP}. 
Then we delete one edge from ${\mathcal T}_1$ and one edge from ${\mathcal T}_2.$ This way we get two paths. We will get a new tour by connecting the  
endpoints of the paths; this has cost at most twice the diameter, which is at most $4 \l/(1-\l).$ The total additional cost is therefore $4 \cdot Top(k)+4\l/(1-\l) \le 8 \cdot Top(k)/(1-\l).$ 
\hfill\qedf

\medskip

It will be convenient to introduce a new notation $L$ for a smooth functional. We will assume that $L(X) \le L(Y)+K\cdot Top(k)/(1-\l),$ if $H(X, Y)\le k,$ with $K>0$ a constant.
We further assume that ${\mathbf E}L=\Theta(\sum_{i=1}^h(b\l)^i).$ 
Notice that by Lemmata~\ref{smoothness} and~\ref{smoothness-TSP} we have the smoothness conditions for $M$, $TSP$ and $MST$, albeit with different values for $K.$  We showed
in previous sections that for these functionals the expectation has the above form. As it turns out, smoothness and
expectation value are 
the most important properties we need for showing strong concentration about the mean. 
 
The isoperimetric inequality we need is standard, but for completeness we present the proof. For a subset $A \subset \Omega$,
we define $$H(A, X) = \min_{Y \in A}H(X, Y).$$ Let us fix a set $A$
such that $\mu^n(A) \ge 1/2$. For a number $t$ we define $$A_t= \{X:
H(X, A) \le t\}.$$ We will show, with the aid of Azuma's inequality, that
$\mu^n(\overline{A_t})=1-\mu^n(A_t)$ tends to zero very fast as we
increase $t.$
Let $$\alpha = \int H(A, X) \ d\mu^n;$$ this is the expected Hamming
distance of a randomly chosen $n$-tuple of $\Omega^n$ from $A.$ Then
by Azuma's inequality, $$\mu^n(\{X: |H(A, X) - \alpha | \ge t\}) \le 2
e^{-{t^2 / 2n}}.$$ (Here we have used the fact that changing one
coordinate results in a change of at most 1 in the Hamming distance.)

Next we give a bound for $\alpha.$ Observe that when $Y \in A$ we have
$H(A, Y) =0.$ Thus $$A \subset \{X : |H(A, X) - C| \ge C\}$$ for all
$C > 0.$ Setting $t=\alpha$ in the above inequality, we have
$$1/2\le \mu^n(A) \le \mu^n(\{X: |H(A, X)-\alpha| \ge \alpha\}) \le 2
e^{-{\alpha^2 / 2n}},$$ which implies $\alpha \le \sqrt{2n\log
  4}.$ Hence $$\mu^n(\{X: H(A, X) \ge t + \sqrt{2n\log 4}\}) \le 2
e^{-{t^2 / 2n}}.$$

\smallskip

Some consideration of the cases $t \ge 2\sqrt{2n\log 4}$ and $t < 2\sqrt{2n\log
  4}$, after necessary modifications of the parameters, produces the
following {\it isoperimetric inequality}, valid for all $t$
independently of $\alpha.$
\begin{proposition}[isoperimetric inequality]\label{iso_ineq}
If $A \subset \Omega^n$ satisfies $\mu^n(A)\ge 1/2,$
then $$\mu^n(\overline{A_t}) \le 4 e^{-{t^2 / 8n}}.$$
\end{proposition}

\medskip

We are going to use the {\it median} of the $L$
functional: $$med(L) = \inf \{t \in {\mathbf R}: \mu^n(\{X \in
\Omega^n: L(X) \le t\})\ge 1/2\}.$$ Let
$$A=\{X \in \Omega^n: L(X) \le med(L)\};$$ then $\mu^n(A) \ge 1/2.$
Applying Lemma~\ref{smoothness}, we obtain
\begin{eqnarray*}
{\mathbf P}(L(X) \ge med(L)+t )&=&\mu^n(\{X \in \Omega^n: L(X) \ge
med(L)+t\}) \\ &\le& \mu^n(\{X \in \Omega^n: med(L)+K\cdot Top(k) /
(1-\lambda) \ge med(L)+t\}) \\ &=& \mu^n(\{X \in \Omega^n: K\cdot
Top(k) / (1 - \lambda) \ge t\}),
\end{eqnarray*}
where $k=H(A, X).$ One can also prove a similar inequality using the
set $B=\{X \in \Omega^n: L(X) \ge med(L)\},$ to bound the
probability ${\mathbf P}(L(X) \le med(L)-t ).$ Putting the two
together, we see

$${\mathbf P}(|L(X) - med(L)|\ge t ) \le 2\mu^n(\{X \in \Omega^n:
K\cdot Top(k) / (1-\lambda) \ge t\}).$$

In the cases that interest us most, we can find the inverse of
$Top(k),$ and hence we can apply the isoperimetric inequality (Proposition \ref{iso_ineq}) to prove
strong concentration around the expected value.

First, we consider the case when $b \lambda =1.$ Then $$Top(k) =
\sum_{i=1}^{\lceil \log_b k \rceil}(b \lambda)^i=\lceil \log_b k
\rceil.$$ In this case, using the isoperimetric inequality, we have

$${\mathbf P}(|L(X) - med(L)|\ge t ) \le 2\mu^n(\{X \in \Omega^n:
KC(\lambda)Top(k) \ge t\})\le $$
$$2\mu^n(\{X \in \Omega^n: k \ge b^{t \over c}\}) \le 4 \exp\left
({-{b^{t/c'} \over 8n}}\right )$$ with the constants $c$ and $c',$
where $c=2c'.$

Now suppose $b\lambda >1.$ Then $$Top(k) = \sum_{i=1}^{\lceil
  \log_b k \rceil}(b \lambda)^i = \frac 1 {1 - \lambda} \cdot b\lambda
{(b\lambda)^{\lceil \log_b k \rceil}-1\over b\lambda -1}\le C_1(b,
\lambda)(b \lambda)^{\log_bk},$$ where $C_1(b, \lambda)$ is a constant
depending only on $b$ and $\lambda$.  The inequality $K \cdot Top(k) /
(1-\lambda) \ge t,$ using that $(b\l)^{\log_b k}=k^{\log_b(b\l)},$ implies the inequality $$k \ge \left({t \over
  C_2(b, \lambda)}\right)^{1 \over \log_b(b\lambda)}.$$ This in turn
implies

$${\mathbf P}(|L(X) - med(L)|\ge t ) \le 2\mu^n(\{X \in \Omega^n:
K\cdot Top(k) / (1-\lambda) \ge t\})\le $$

$$2\mu^n(\{X \in \Omega^n: k \ge (t /C_2(b, \lambda))^{1 \over
  \log_b(b\lambda)}\}) \le 4 \exp\left ({-{t^{2/\log_b (b\lambda)}
    \over C_3 n}}\right )$$ for $b\lambda >1.$
 
Since $\int {\mathbf P}(Z \ge t) \ dt = {\mathbf E}Z$ for any random
variable $Z,$ integrating these inequalities over the non-negative
reals produces an upper bound for ${\mathbf E} |L(X)-med(L)| \ge
|{\mathbf E} L(X) - med(L)|.$ Hence, we can derive concentration
inequalities for the probability ${\mathbf P}(|L(X)-{\mathbf E}L(X)| \ge t).$
In the next section we will look at some important special cases.

\subsection{Important special cases}

It is particularly interesting to consider the case $\lambda = 1/2,$
since it relates the functionals in question on an HST with functionals on the
unit cube of some dimension. When $b=2^d$ for some positive integer $d,$
the HST approximates the $d$-dimensional unit cube in Euclidean space.

Let us first consider the case $b=2$:  we encounter such a tree $T$ when
approximating the $[0,1]$ interval. For this case $b \lambda =1,$ and
therefore $Top(k) = O(\log_2 k).$ It is easy to see (using, e.g.,
numerical approximation) that 
$$|{\mathbf E} L(X) - med(L)|\le \int_{t\ge 0} {\mathbf P}(|L(X) - med(l)|\ge t ) \ dt \le 4 \int_{t \ge 0} \exp\left ({-{2^{t/c} \over 8n}}\right ) \ dt =
\Theta(\log_2 n).$$

Consequently, we have 
$${\mathbf P}(|L(X)-{\mathbf E}L(X)| \ge t+C\log_2 n) \le 4 \exp\left(-{2^{t/c}\over 8n}\right).$$
Since ${\mathbf E}L(X) = O(\log_2 n),$ this means that we don't have an especially useful concentration
result for this special case.  Inspecting cases, one finds the following
inequality for the case $b=2$ and $\lambda=1/2$:
$${\mathbf P}(|L(X)-{\mathbf E}L(X)| \ge t) \le 8 \exp\left(-{2^{t/c}\over 8n}\right),$$
where $c$ is a real constant. While the constants can be
improved somewhat (partly due to the fact that the constants
are not the best possible for this case), this
inequality allows $L(X)$ to be spread out over an interval of length
$\Theta({\mathbf E}L(X)).$ However, the probability of $L(X)$ falling
outside this interval is very small: even a deviation of $C\log \log
n$ results in a probability less than $1/n.$ 



On the other hand, the bounds that we have derived will provide very
good concentration inequalities for the case $b > 2.$ As above, we need to estimate an integral to get a bound for
$ |{\mathbf E} L(X) - med(L)|:$ 
$$|{\mathbf E} L(X) - med(L)|\le 4\int_{t\ge 0} \exp\left ({-{t^{2/\log_b (b\lambda)} \over C_3 n}}\right )\ dt=\Theta(\sqrt{{\mathbf E}L(X)}),$$ as can be
shown by numerical approximation.

We thus have the following concentration inequality:
$${\mathbf P}(|L(X)-{\mathbf E}L(X)| \ge t) \le 8 \exp\left
({-{t^{2/\log_b (b/2)} \over c'n}}\right ),$$ where $c'$ is a real
constant. Notice that $\log_b(b/2) <1,$ hence, whenever $\lambda=1/2$ and $b\ge 3,$ the exponent of $t$ above is
larger than $2.$  In other words, in these cases all the considered monochromatic optimization
problems exhibit sub-Gaussian behavior.

The case $b=2^d$ can be used to approximate the $L$ functional in the $d$-dimensional unit cube. In general, when $b=2^d$ the 
inequality has the following form:
$${\mathbf P}(|L(X)-{\mathbf E}L(X)| \ge t) \le 8
\exp\left({-{t^{2d/(d-1)} \over c''n}}\right),$$ for some real
constant $c''.$ 


\subsection{The bichromatic case}
 
 In this case we lack a counterpart to
 Lemma~\ref{smoothness}. What we have instead is that a change at one
 point changes the length of the red-blue matching by at most one unit
 (the diameter of $T$ is a constant, and after normalization it is one
 unit).  We can apply Azuma's inequality, giving us
$${\mathbf P}(|M(X)-{\mathbf E}M(X)| \ge t) \le 2 \exp \left ({-{t^2 \over 2n}}\right ).$$ 
This is clearly weaker than the very strong concentration results we obtained for
the monochromatic matching problems. In general we cannot expect any better than that.
Let's consider the star tree with two leaves. The difference of the red and blue points at one of the leaves
is approximately normal if $n$ is large, since the red and the blue points have approximately normal distribution
when $n$ is large. This implies that the discrepancy cannot have a sub-gaussian behavior.

\section{Conclusion}\label{conclusion}


In this paper, we considered five optimization problems --- monochromatic matching, and monochromatic and bichromatic
versions of the MST problem and the TSP --- on points randomly distributed among the leaves of a balanced hierarchically 
separated tree.  These investigations continued the work of Abrahamson \cite{AbrahamsonPhD}
and his collaboration with Csaba and Shokoufandeh \cite{Abrahamson}, in which
the bichromatic matching problem was considered.
We were able to come up with tight upper and lower bounds to these five problems that
resemble those obtained in \cite{AbrahamsonPhD} and \cite{Abrahamson} for bichromatic
matching.
We also obtained concentration inequalities for the monochromatic problems;
in passing, we noted that the bichromatic problems present substantial difficulty.

It will be useful to extend these results to more general metric spaces than we have considered here.
For instance,
Bartal's original definition of HST \cite{Bartal96} did not enforce an exact ratio of $\lambda$ between parent
and child edges, but rather that the ratio was at most $\lambda$.  One can also try to embed arbitrary metric spaces
into an HST, turning optimization problems in continuous sets into HST optimization problems.  Here, of course,
we must account for some distortion.  These applications constitute a great part of the utility of HSTs,
but we leave their consideration for future work.


\begin{thebibliography}{50}

\bibitem{AbrahamsonPhD} Jeff Abrahamson. {\it Optimal Matching and Determininstic Sampling.} PhD thesis, Drexel University, 2007.

\bibitem{Abrahamson} Jeff Abrahamson, B\'ela Csaba, and Ali Shokoufandeh. Optimal random matchings on trees and applications. In {\it Proceedings of the 12th International Workshop on Randomization and Computation,} RANDOM '08, pages 254-265, Berlin, Heidelberg, 2008. Springer-Verlag.

\bibitem{Ajtai84} M. Ajtai, J. Koml\'os, and G. Tusn\'ady. On optimal matchings. {\it Combinatorica,} 4(4):259-264, 1984.


\bibitem{Bartal96} Yair Bartal. Probabilistic approximation of metric spaces and its algorithmic applications. In {\it Proceedings of the 37th Annual Symposium on Foundations of Computer Science,} pages 184-193, 1996.


\bibitem{Bartal98} On approximating arbitrary metrics by tree metrics. In {\it Proceedings of 30th Annual ACM Symposium on Theory of Computing,} pages 161-168, 1998.


\bibitem{Fakcharoenphol03} Jittat Fakcharoenphol, Satish Rao, and Kunal Talwar. A tight bound on approximating arbitrary metrics by tree metrics. In {\it Proceedings of the 35th annual ACM Symposium on Theory of Computing,} STOC '03, pages 448-455, New York, NY, USA, 2003. ACM.

\bibitem{Indyk} Piotr Indyk and Nitin Thaper. Fast image retrieval via embeddings. In {\it 3rd International Workshop on Statistical and Computational Theories of Vision (at ICCV),} 2003.


\bibitem{Leighton86} F. T. Leighton and P. Shor, Tight bounds for minimax grid matchings, with applications to the average case analysis of algorithms. In {Proceedings of the 18th annual ACM Symposium on Theory of Computing,} STOC'86, pages 91-103, New York, NY, USA, 1986. ACM.

\bibitem{Rhee92} Wansoo T. Rhee. On the travelling salesperson problem in many dimensions. {\it Random Structures and Algorithms,} 3(3): 227-233, 1992.

\bibitem{Rhee93} Wansoo T. Rhee. A matcing problem and subbadditive Euclidean functionals. {\it The Annals of Applied Probability,} 3(3): 794-801,1993.


\bibitem{Rhee88} Wansoo T. Rhee and Michel Talagrand. Exact bounds for the stochastic upward matching problem. {\it Transactions of the American Mathematical Society,} 109(1): 109-125, 1988.

\bibitem{Rubner00} Yossi Rubner, Carlo Tomasi and Leonidas J.~Guibas. The earth mover's distance 
as a metric for image retrieval. {\it International Journal of Computer Vision,} 40:99-122, 2000.

\bibitem{Shor86} The average-case analysis of some on-line algorithms for bin packing. In {\it FOCS 1984: \\
25th Annual Symposium on Foundations of Computer Science,} pages 193-200, 1984.

\bibitem{Steel97} J. Michael Steele. {\it Probability Theory and Combinatorial Optimization.} SIAM, 1997.


\bibitem{Talagrand92} Michel Talagrand. Matching random samples in many dimensions. {\it The Annals of
Applied Probability,} 2(4):846-856, 1992. 

\bibitem{Talagrand95} Michel Talagrand. Concentration of measure and isoperimetric inequalities in product spaces. {\it Publications math\'ematiques de l'I.H.E.S.,} 81:73-205, 1995.

\bibitem{Talagrand96} Michel Talagrand. Majorizing Measures: the Generic Chaining. {\it The Annal of Probability,} 24(3): 1049-1103, 1996.

\bibitem{Yukich} Joseph E. Yukich. {\it Probability Theory of Classical Euclidean Optimization Problems.} Springer, 1998.

\end{thebibliography}
\end{document}